\documentclass[conference]{IEEEtran}
\usepackage{amsmath,amssymb}
\usepackage{graphicx}
\usepackage[font=small,format=plain,labelfont={bf,up},labelsep=space]{caption}

\DeclareCaptionLabelFormat{table}{TABLE~#2}
\usepackage{caption}
\usepackage{subcaption}
\usepackage{amsthm}
\newtheorem{theorem}{Theorem}
\newtheorem{property}{Property}
\usepackage{amsmath}
\usepackage{lipsum}
\usepackage{float}
\usepackage{multicol}
\usepackage{longtable}
\usepackage{comment}
\newtheorem{illustration}{Illustration}

\usepackage{caption}
\usepackage{subcaption}
\usepackage{float}

\ifCLASSINFOpdf
\else
\fi
\begin{document}
%
\title{Multi-Number CVT-XOR Arithmetic Operations In Any  Base System And Its Significant  Properties}

\author{\IEEEauthorblockN{Jayanta Kumar Das}
\IEEEauthorblockA{Applied Statistics Unit \\ Indian Statistical Institute\\
Kolkata-700108, India\\
Email:dasjayantakumar89@gmail.com}
\and
\IEEEauthorblockN{Pabitra Pal Choudhury}
\IEEEauthorblockA{Applied Statistics Unit \\ Indian Statistical Institute\\
Kolkata-700108, India\\
Email:pabitrapalchoudhury@gmail.com}
\and
\IEEEauthorblockN{Sudhakar Sahoo}
\IEEEauthorblockA{Computer Science\\ Institute of Mathematics and Applications\\ Bhubaneswar-751003, India\\
Email: sudhakar.sahoo@gmail.com}
}

%


\maketitle

\begin{abstract}
Carry Value Transformation (CVT) is a model of discrete dynamical system which is one special case of Integral Value Transformations (IVTs). Earlier in [5] it has been proved that sum of two non-negative integers is equal to the sum of their $CVT$ and $XOR$ values in any base system. In the present study, this phenomenon is extended to perform $CVT$ and $XOR$ operations for many non-negative integers in any base system. To achieve that both the definition of $CVT$ and $XOR$ are modified over the set of multiple integers instead of two. Also some important properties of these operations have been studied. With the help of cellular automata the adder circuit designed in $[14]$ on using $CVT-XOR$ recurrence formula is used to design a parallel adder circuit for multiple numbers in binary number system.

Keywords- Integral Value Transformations; Carry Value Transformation; Recursion, Adder circuit etc. 
\end{abstract}


%
\IEEEpeerreviewmaketitle

\section{Introduction}
Integral Value Transformations (IVTs) is a class of continuous maps in a discrete space and was introduced $[1, 2, 7]$ in the year 2009.
A p-adic, k-dimensional, Integral Value Transformation is denoted by $IVT^{p,k}_{j}$ and it is a mapping from $N_{0}^{k} \longrightarrow N_{0}$. When k=1, $IVT^{p,1}_{j}$ is defined as\\
$
\begin{array}[b]{lcr}
     IVT^{p,1}_{j}(x)=   
     ( 
          f_{j}(x_{n})f_{j}(x_{n-1})...f_{j}(x_{1}))_{p}=(m)_{10}
 \end{array}$
 
where $m$ is the decimal conversion from the p-adic number and $x$ is non-negative p-adic integer represented as $x=(x_{n} x_{n-1}...x_{1})_{p}$ and the Rule number denoted by $f_{i}$ is a local mapping defined from $\{ 0, 1, 2, 3,...,p-1\}$ to $\{ 0, 1, 2, 3,...,p-1\}$. Here $j$ is the decimal equivalent of the p-adic string in the truth table representation of the local map.
For example, when $p=3$, $k=1$ and say $x=(14)_{10}=(112)_{3}$ and for the two different Rule numbers $5$ and $16$ shown in TABLE I, the $IVTs$ are calculated as
$IVT^{3,1}_{5}(14)=(f_{5}(1)f_{5}(1)f_{5}(2))_{3}=(110)_{3}=(12)_{10}$ and 
$IVT^{3,1}_{16}(14)=(f_{16}(1)f_{16}(1)f_{16}(2))_{3}=(221)_{3}=(25)_{10}$ 

\begin{table}[ht]
\caption{Truth Table of two 1-variable ternary functions (base 3) functions $f_{5}$ and $f_{16}$}
\centering 
\begin{tabular}{|c|c|c|} 
\hline
\bf Variable & \multicolumn{2}{|c|}{\bf Rule} \\
\hline
\bf $x_i$ & $\bf {f_{5}}$& $\bf {f_{16}}$\\
\hline
0 & 2 & 1 \\
\hline
1 & 1 & 2 \\
\hline
2 & 0 & 1 \\
\hline
\end{tabular}
\label{table:nonlin} 
\end{table}

Like One dimensional, two dimensional p-adic, Rule $j$ $IVT$ denoted by $IVT_{j}^{p,2}(x,y)$ is defined as $IVT^{p,2}_{j}(x,y)=(f_{j}(x_{n},y_{n})f_{j}(x_{n-1},y_{n-1})...f_{j}(x_{1},y_{1}))_{p}=(m)_{10}$. Similarly, k-dimensional $IVTs$ can be defined. (Sometimes the symbol $\beta$ is used instead of $p$ as base of the number system)

Carry Value Transformation (CVT) which was initially defined in the year 2008 $[3]$, later developed and elaborated in $[4, 5]$ became a special case of $IVT_{j}^{p,k}(x, y)$ when $p=2$, $k=2$ and $j=8$ along with a $0$ padded in the LSB position of the output binary string. Thus $CVT$ is a two dimensional, Rule $8$, binary $IVT$ with a $0$ padded in it where as $XOR$ is simply a two dimensional Rule $6$ binary $IVT$ as shown in TABLE II. 
\begin{table}[ht]
\caption{Truth Table of two 2-variable Boolean functions $f_{8}$ and $f_{6}$}
\centering 
\begin{tabular}{|c|c|c|} 
\hline
\bf Variable & \multicolumn{2}{|c|}{\bf Rule} \\
\hline
\bf $x_i$ $y_{i}$ & $\bf {f_{8}}$& $\bf {f_{6}}$\\
\hline
0 0 & 0 & 0 \\
\hline
0 1 & 0 & 1 \\
\hline
1 0 & 0 & 1 \\
\hline
1 1 & 1 & 0 \\
\hline
\end{tabular}
\label{table:nonlin} 
\end{table}

Carry Value Transformations were studied in $[3]$ to produce self-similar fractal whose dimension is same as the dimension of the Sierpinski triangle. Further they have shown that $CVT$ can also be used to produce many periodic and chaotic patterns. Also, analytical and algebraic properties of $CVT$ were studied in $[3]$.  Different fractals having dimension in between $1$ and $2$ were studied in $[4]$. Two most important properties of $CVT$ and Modified Carry Value Transformation (MCVT) were studied in $[5]$. Where It has been shown that $(1)$ sum of two non-negative integers are equal to their $CVT$ and $XOR$ values i.e. $a+b=CVT(a,b)+XOR(a,b)$ in any number system and  $(2)$ the number of iterations leading to either $CVT=0$ or $XOR=0$ does not exceed the maximum of the lengths of the two addenda expressed as binary strings i.e. the convergence behaviours of $CVT$ and $MCVT$ were discussed. Some similar kind of transformations such as Extreme Value Transformation (EVT) $[4]$, $2-$ Variable Boolean Operation (2-VBO) $[5]$, Integral Value Transformations (IVTs) $[7]$ are also used to manipulate strings of bits and applicable in pattern formations $[4, 7]$, solving Round Rabin Tournament problem $[8]$, Collatz-like functions $[7]$ and so forth. Previously used adder circuits $[9, 10, 11, 12]$  are combinational in nature and their complexity depends on number of logic gates used and the associated gate delays. In line with this Cellular Automata Machines $[13]$, were studied in $[14]$ for efficient hardware design of some basic arithmetic operations where their complexity centered on number of clock cycles required to finish the computation instead of the gate delays.

The organization of this paper is as follows: In section II some of the preliminary concepts on $CVT$ and $XOR$ operation of two numbers in binary domain is highlighted (also, thoroughly elaborated in $[3]$). In section III we have discussed the $CVT$ and $XOR$ operations of many numbers in any base system and studied some of its important properties. In section IV a parallel architecture for multi number addition in binary number system has been proposed. In Section V conclusion for this article along with some future research planning have been added.

\section{Carry Value Transformation}
The carry or overflow bits are usually generated at the time of addition between two $n$-bit strings. In the usual addition process, carry value is always a single bit and if generated then it is added column wise with other bits and not necessarily save for further use. But the carry value defined in $[3]$ are the usual carries generated bit wise and stored in their respective places as shown in TABLE III.

\begin{table}[ht]
\caption{Carry generated in the $i^{th}$ column counted from LSB is saved in $(i+1)^{th}$ column.} 
\centering 
\resizebox{8cm}{!} {
\begin{tabular}{c c c c c c c c} 
carry value & = & $a_{n} \wedge b_{n}$& $a_{n-1} \wedge b_{n-1}$ &$a_{n-2} \wedge b_{n-2}$&...&$a_{1} \wedge b_{1}$& 0 \\ 
a &=  & & $a_{n}$ & $a_{n-1}$&...& &$a_{1}$ \\
b &= & & $b_{n}$ & $b_{n-1}$&...& &$b_{1}$ \\
\hline
$a \oplus b$&=& & $a_{n} \oplus b_{n}$&$a_{n-1} \oplus b_{n-1}$ &...& &$a_{1} \oplus b_{1}$ \\
\hline 
\end{tabular}
}
\label{table:nonlin} 
\end{table}
  
Thus to find out the carry value we perform the bit wise $XOR$ operation of the operands to get a string of sum-bits (ignoring the carry-in) and simultaneously the bit wise ANDing of the operands to get a string of carry-bits, the latter string is padded with a $0$ on the right to signify that there is no carry-in to the LSB. Thus the corresponding decimal value of the string of carry bits is always an even integer. 
Precise form of $CVT$ is a mapping defined as $(B_{n}\times B_{n}) $to$ B_{n} $ where  $B_{n}$ is the set of strings of length $n$ on $B=\{0,1\}$ . More specifically, if  $a=(a_{n},a_{n-1},...,a_{1})$ and  $b=(b_{n},b_{n-1},...,b_{1})$ then  $CVT(a,b)=(c_{n},c_{n-1},...,c_{1},0)$ where  $c_{n}=a_{n}\wedge b_{n}$ is an $n+1$  bit strings, belonging to set of non-negative integers and can be computed bit wise by logical AND operation followed by a $0$ which denotes no carry is generated in the LSB at the time of addition procedure.\\

\begin{illustration}
Suppose we want to find out the $CVT$ of two non-negative integers say 11 and 14. First of all, we have to find out the binary representation of these numbers . Both are 4-bits numbers. The carry value is calculated as shown in TABLE IV:
\begin{table}[ht]
\caption{Carry generated in $ith$ column is saved in $(i+1)th$ column for $i=0, 1, 2,3,4$.
} 
\centering 
\begin{tabular}{c c c c c c c} 
Carry& : & 1& 0 &$ 0$& 1& 0 \\ 
\hline
a &: & & 1 & 0 & 1 & 1 \\
b &: & & 1 & 1 & 0 & 1 \\
\hline
XOR &: & & 0 & 1 & 1 & 0 \\
\hline 
\end{tabular}
\label{table:nonlin} 
\end{table}
\end{illustration}

Conceptually, in the general addition process the carry or overflow bit from each stage (if any) goes to the next stage so that, in each stage after the first (i.e. the LSB position with no carry-in), actually a 3-bit addition is performed instead of a $2-$bit addition by means of the full adder. Instead of going for this traditional method, what we do is that we perform the bit wise $XOR$ operation of the operands (ignoring the carry-in of each stage from the previous stage) and simultaneously the bit wise ANDing of the operands to get a string of carry-bits, the latter string is padded with a $0$ on the right to signify that there is no carry-in to the LSB (the overflow bit of this ANDing being always $0$ is simply ignored). In our example, bit wise $XOR$ gives $(0110)_{2}=(6)_{10}$  and bit wise ANDing followed by zero-padding gives $(10010)_{2}=(18)_{10}$. Thus   $CVT(1011,1101)_{2}=(10010)_{2}$ and equivalently in decimal notation one can write $CVT(11,13)=18$. Figure 1 shows the circuit diagram of CAM used for performing addition of two 4-bit numbers $[14]$. This CAM is based on a recurrence relation
\begin{equation}
X+Y=CVT(X, Y)+XOR(X,Y)
\end{equation}
 which has been proved to be valid for any base system [5].
\begin{figure}[h]
\begin{center}
\includegraphics[scale=.65]{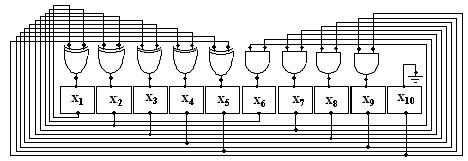}
\caption{Figure shows the CAM design for 4-bit addition circuit}
\end{center}
\label{fig:1}
\end{figure}

This CAM is used to design an adder circuit for multiple numbers in binary number system and proposed in section IV.
\section{CVT and XOR Operations of Many Numbers In Any Base System}
\textbf{Definition:}
For any number system in base $\beta,$  $XOR$ and $CVT$ of $K$ non-negative integers is defined as follows: (here $K$ integers represented as $X_{1}, X_{2}, X_{3},...,X_{K}$)
$
\begin{array}{lcr}
      X_{1}=a_{1n},...,a_{11}\\
      X_{2}=a_{2n},...,a_{21}\\
      .\\
      .\\
      .\\
      X_{K}=a_{kn},...,a_{k1}\\
      \hline
      
      XOR(X_{1}, X_{2},...,X_{k})=((a_{1n}+a_{2n}+...+a_{Kn}) \mod\\ \beta,...,(a_{11}+a_{21}+,...,+a_{K1}) \mod \beta)_\beta=R=(R_{n}, R_{n-1}, ..., R_{1})_{\beta}$ 
     $ \end{array}$ \\ 
      $     
      \begin{array}{lcr}
     
      $ and $CVT(X_{1},X_{2},X_{3},...,X_{K})=C=(C_{n},C_{n-1},...,C_{1},0)_{\beta}$
     $\\$ where $C_{i}=\lfloor (a_{1i}+a_{2i}+,...,+a_{Ki})/\beta \rfloor$
      and $i=1, 2, 3,...,n.
\end{array}$ \\
Note: In general, $C_{i}$ result may not be in base $\beta$ system but the decimal conversion of $(C_{n},C_{n-1},...,C_{1},0)=(C_{n}\times \beta^{n-1}+C_{n-1}\times \beta^{n-2}+...+C_{1}\times \beta^{1}+0)$ is same as the method from base $\beta$ to base 10.
\begin{theorem}
The recurrence relation in equation (1) is also valid for many numbers in any base system.\\
That is $X_{1}+X_{2}+X_{3}+...+X_{K}=CVT(X_{1}, X_{2}, X_{3},...,X_{K})+XOR(X_{1}, X_{2}, X_{3},...,X_{K})$
\end{theorem}
The proof of the above theorem can be similarly seen by extending the proof in [5].

\begin{illustration}
Suppose in ternary number system (i.e. $\beta=3)$ $CVT$ and $XOR$ operation of decimal numbers $17, 8, 11, 8, 4, 8$ are as follows:\\ \\
$CVT(17, 8, 11, 8, 4, 8)=CVT(122,022, 102,022, 011, 022)\\=(\lfloor 2/3 \rfloor \lfloor 9/3 \rfloor \lfloor 11/3 \rfloor 0)=(0330)=(0 \times 3^{3}+3 \times 3^{2}+3 \times 3^{1}+0 \times 3^{0})=(36)_{10}$ and\\ \\
$XOR(17, 8, 11, 8, 4, 8)=XOR(122,022, 102,022, 011, 022)=(202)_{3}=(20)_{10}$.\\
So we have observed that $17+8+11+8+4+8=CVT(17, 8, 11, 8, 4, 8)+XOR(17, 8, 11, 8, 4, 8)=36+20$. 
\end{illustration}

\subsection{Important Corollaries}
Following two important Corollaries i.e. Corollary 1 and Corollary 2 are trivially obtained from the definitions of CVT and XOR operations for any base system and for arbitrary $K$ numbers.

{\bf Corollary 1.} 
From XOR definition, $R_{i}=\{0,1,2,...,\beta-1\}$ and 
$R_{i}=0$ iff $R_{i}$ is a multiple of $\beta$. So $XOR(X_{1}, X_{2},...,X_K$)=0 iff $R_{i}$ is multiples of $\beta$ $\forall i$.  \\
 \\
{\bf Corollary 2.} 
Similarly from CVT definition, $C_{i}=\{ 0,1,2...\frac{k\times (\beta -1)}{\beta}\}$ and
 $C_{i}=0$ iff $C_{i}<\beta$. 
 So $CVT(X_{1}, X_{2},...,X_{K}$)=0 iff $C_{i}<\beta $ $\forall i$.

\subsection{Important Properties of Multi Numbers CVT and XOR Operations in Binary Number System}
\subsubsection{CVT-Property}

\begin{property}
For base $\beta$ system, if all the $K$ numbers are same then

\begin{enumerate}
  \item[(a)] if $K$ is even, $CVT$ is equal to addition of $K$ numbers i.e.
   $CVT (X, X,...,{K}$ times)$ = X + X + ...+K$ times= $K \times X$.
   \item[(b)] if $K$ is odd, $CVT$ is equal to addition of  $(K-1)$ numbers i.e.
   $CVT(X, X,...K$ times)$ = X + X + ...+(K-1)$ times=$ (K-1) \times X$.
\end{enumerate}
\begin{illustration}
For base $\beta=2$ and take X=5 (101)\\
if K is even (say K=4) then \\
$CVT(5, 5, 5, 5)=CVT(101, 101, 101, 101)=(2020)_{2}=(20)_{10}=4\times 5=K\times X$\\
if K is odd (say K=5) then\\
$CVT(5, 5, 5, 5, 5)=CVT(101, 101, 101, 101, 101)=(2020)_{2}=(20)_{10}=(5-1)\times 5=(K-1)\times X$.
\end{illustration}
\end{property}
 
\begin{property}
For base $\beta$ system, if $CVT(X_{1},X_{2},...,X_{n})=P$ and let $K$ is a scalar quantity and is a power of $\beta$ then
\begin{enumerate}
\item[(a)] $CVT(K\times X_{1},K\times X_{2},...,K\times X_{n})=K\times P$ and 
\item[(b)] $CVT(\frac{X_{1}}{K},\frac{X_{2}}{K},...,\frac{X_{n}}{K} )=\lfloor\frac{P}{K+m}\rfloor$; where  $m=\frac{\text{number of odd numbers}}{\frac{K}{2}}$
\end{enumerate}
\begin{illustration}
For base $\beta=2$ and let $X_{1}=5 (101), X_{2}=4 (100), X_{3}=6 (110), X_{4}=7 (111)$ and $K=4$. \\ 
$CVT(5,4,6,7)=$ $CVT(101, 100, 110, 111)=(2110)_{2}=(22)_{10}=P$ then \\ 
$CVT(4\times 5,4\times 4,4\times 6,4\times 7)=CVT(10100, 10000, 11000, 11100)=(211000)_{2}=(88)_{10}=4\times 22=K\times P$ and \\
$CVT(\frac{5}{4}, \frac{4}{4}, \frac{6}{4}, \frac{7}{4})=CVT(1, 1, 1, 1)=CVT(1, 1, 1, 1)=(20)_{2}=(4)_{10}=\lfloor\frac{22}{4+1}\rfloor=\lfloor\frac{P}{K+m}\rfloor$.
\end{illustration}
\end{property}

\begin{property}
If $CVT(X_{1}, X_{2},...,X_{n})=P$ and \\ $CVT(Y_{1}, Y_{2},...,Y_{n})=Q$ then \\
$CVT(X_{1}, X_{2},...,X_{n},Y_{1}, Y_{2},...,Y_{n})=P+Q$ 
\begin{illustration}
For base $\beta=2$ and let $X_{1}=5 (101), X_{2}=4 (100), X_{3}=6 (110), X_{4}=7 (111)$ and $Y_{1}=13 (1101), Y_{2}=9 (1001), Y_{3}=9 (1001), Y_{4}=13 (1101)$. \\ 
$CVT(5,4,6,7)=$ $CVT(101, 100, 110, 111)=(2110)_{2}=(22)_{10}=P$ and \\ 
$CVT(13,9,9,13)=$ $CVT(1101, 1001, 1001, 1101)=(21020)_{2}=(44)_{10}=Q$  then\\
$CVT(5,4,6,7,13,9,9,13)=$ $CVT(0101, 0100, 0110, 0111, 1101, 1001, 1001, 1101)=(23130)_{2}=(66)_{10}=P+Q$.
\end{illustration}
\end{property}

\begin{property}
If $CVT(X,X,...,X)=P$ then $CVT(X^{K}, X^{K},...,X^{K})= P\times X^{k-1} $.
\begin{illustration}
For base $\beta=2$ and let $X=3 (11)$, K=3
$CVT(3,3,3,3)=$ $CVT(11, 11, 11, 11)=(220)_{2}=(12)_{10}=P$ then\\ 
$CVT(3^{3},3^{3},3^{3},3^{3})=$ $CVT(11011, 11011, 11011, 11011)=(220220)_{2}=(108)_{10}=12\times 3^{3-1}=P\times X^{K-1}$.
\end{illustration}
\end{property}

\subsubsection{XOR-Property}
\begin{property}
If all the $K$ numbers are same then
 \begin{enumerate}
\item[(a)] if $K$ is even, $XOR$ of $K$ numbers  is zero i.e. $XOR (X, X,...K$ times) $= 0$.
\item[(b)] if $K$ is odd, $XOR$ of $K$ numbers is equal to a single number i.e. $XOR (X, X,...K$ times) $= X$.
 \end{enumerate}
 \begin{illustration}
 For base $\beta=2$ and take X=5 (101)\\
 if K is even (say K=4) then \\
 $XOR(5, 5, 5, 5)=XOR(101, 101, 101, 101)=(0)_{2}=0$\\
 if K is odd (say K=5) then\\
 $XOR(5, 5, 5, 5, 5)=XOR(101, 101, 101, 101, 101)=(101)_{2}=(5)_{10}=X$.
 \end{illustration}
\end{property}

\begin{property}
For base $\beta$ system, if $XOR(X_{1},X_{2},...,X_{n})=Q$ and let $K$ is a scalar quantity and is a power of $\beta$ then 
\begin{enumerate}
\item[(a)] $XOR(K\times X_{1},K\times X_{2},...,K\times X_{n})=K\times Q$ and
\item[(b)] $XOR(\frac{X_{1}}{K},\frac{X_{2}}{K},...,\frac{X_{n}}{K} )=\lfloor\frac{Q}{K}\rfloor$.
\end{enumerate}
\begin{illustration}
For base $\beta=2$ and let $X_{1}=5 (101), X_{2}=4 (100), X_{3}=5 (101), X_{4}=7 (111)$ and $K=4$. \\ 
$XOR(5,4,5,7)=$ $XOR(101, 100, 101, 111)=(011)_{2}=(3)_{11}=Q$ then \\ 
$XOR(4\times 5,4\times 4,4\times 5,4\times 7)=XOR(10100, 10000, 10100, 11100)=(01100)_{2}=(12)_{10}=4\times 3=K\times Q$ and \\
$XOR(\frac{5}{4}, \frac{4}{4}, \frac{5}{4}, \frac{7}{4})=XOR(1, 1, 1, 1)=XOR(1, 1, 1, 1)=(0)_{2}=(0)_{10}=\lfloor\frac{3}{4}\rfloor=\lfloor\frac{Q}{K}\rfloor$.
\end{illustration}
\end{property}

\begin{property}
If $XOR(X_{1}, X_{2},...,X_{n})=P$ and $XOR(Y_{1}, Y_{2},...,Y_{n})=Q$ then \\
$XOR(X_{1}, X_{2},...,X_{n},Y_{1}, Y_{n},...,Y_{n})=P\oplus Q$ 
\begin{illustration}
For base $\beta=2$ and let $X_{1}=5 (101), X_{2}=4 (100), X_{3}=5 (101), X_{4}=7 (111)$ and $Y_{1}=13 (1101), Y_{2}=9 (1001), Y_{3}=9 (1001), Y_{4}=10 (1010)$. \\ 
$XOR(5,4,5,7)=$ $XOR(101, 100, 101, 111)=(011)_{2}=(3)_{10}=P$ and \\ 
$XOR(13,9,9,10)=$ $XOR(1101, 1001, 1001, 1010)=(0111)_{2}=(7)_{10}=Q$  then\\
$XOR(5,4,5,7,13,9,9,10)=$ $XOR(0101, 0100, 0101, 0111, 1101, 1001, 1001, 1010)=(0100)_{2}=(4)_{10}=3\oplus 7=P\oplus Q$.
\end{illustration}
\end{property}

\begin{property}
If $XOR(X,X,...,X)=P$ then $XOR(X^{K}, X^{K},...,X^{K})=P\times X^{K-1}$.
\begin{illustration}
For base $\beta=2$ and let $X=3 (11)$, K=2
$XOR(3,3,3)=$ $XOR(11, 11, 11)=(11)_{2}=(3)_{10}=P$ then\\ 
$XOR(3^{2},3^{2},3^{2})=$ $XOR(1001, 1001, 1001)=(1001)_{2}=(9)_{10}=3\times 3^{2-1}=P\times X^{K-1}$.
\end{illustration}
\end{property}

\section{Proposed adder circuit for multiple numbers in binary number system}
The Figure 2 shows the circuit design for $K(=16)$ $4$-bit numbers using CAM. Where $K-1$ CAMs are required. For each CAM internal circuit design is throughly elaborated in $[14]$ and also shown in Figure 1. Initially in  first level, computation is performed on 8 CAMs for two numbers pairwise $(x_{1}, x_{2}) , (x_{3}, x_{4}),...,(x_{15}, x_{16})$ in parallel. Output from each $8$ CAMs are forwarded to the $4$ second level CAMs and so on. Delay in each level is $4$. So maximum delay is $=4\times 4=16$ unit.
\begin{figure}[h]
\begin{center}
\includegraphics[scale=.65]{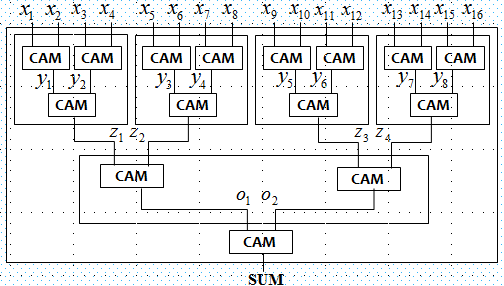}
\caption{Circuit design using CAMs for performing addition of sixteen 4-bit numbers in parallel}
\end{center}
\label{fig:1}
\end{figure}

\section*{Acknowledgment}
The authors would like to give anonymous thank to Dr. Sk Sarif Hassan for providing valuable suggestions to work with this domain.

\section{Conclusion}
Here we have seen how to perform the Multi-Number $CVT$ and $XOR$ Operation in any base system. Some important properties of these operations are highlighted both in any base and binary number system. The implementations of this multi number arithmetic operations in binary system using parallel adder circuit has been proposed. In this context another parallel adder circuit design can be performed on using recurrence relation where lesser number of CAMs are required compare to the circuit design shown in Figure 2.



%

\end{document}